\documentclass[final, 5p, times, sort&compress]{elsarticle}
\usepackage{hyperref}
\usepackage{subcaption}
\usepackage{xcolor}
\journal{arxiv.org}

\begin{document}
\begin{frontmatter}


\title{Development of a large-mass, low-threshold detector system with simultaneous measurements of athermal phonons and scintillation light}


\author[mysecondaryaddress1,mysecondaryaddress2]{M. Chaudhuri}
\author[mymainaddress]{G. Agnolet}
\author[mysecondaryaddress1,mysecondaryaddress2]{V. Iyer}
\author[mysecondaryaddress1,mysecondaryaddress2]{V. K. S. Kashyap}
\author[mymainaddress]{M. Lee}
\author[mymainaddress]{R. Mahapatra}
\author[mymainaddress]{S. Maludze}
\author[mymainaddress]{N. Mirabolfathi}
\author[mysecondaryaddress1,mysecondaryaddress2]{B. Mohanty}
\author[mymainaddress]{M. Platt}
\author[mysecondaryaddress1,mysecondaryaddress2]{A. Upadhyay}
\author[mymainaddress]{S. Sahoo}
\author[mymainaddress]{S. Verma}

\address[mysecondaryaddress1]{School of Physical Sciences, National Institute of Science Education and Research, Jatni 752050, India}
\address[mysecondaryaddress2]{Homi Bhabha National Institute, Training School Complex, Anushaktinagar, Mumbai 400094, India}
\address[mymainaddress]{Department of Physics \& Astronomy, Texas A\&M University, College Station, TX 77843, USA}

\begin{abstract}

  We have combined two low-threshold detector technologies to develop a large-mass, low-threshold detector system that simultaneously measures the athermal phonons in a sapphire detector while an adjacent silicon high-voltage detector detects the scintillation light from the sapphire detector. This detector system could provide event-by-event discrimination between electron and nuclear events due to the difference in their scintillation light yield. While such systems with simultaneous phonon and light detection have been demonstrated earlier with smaller detectors, our system is designed to provide a large detector mass with high amplification for the limited scintillation light. Future work will focus on at least an order of magnitude improvement in the light collection efficiency by having a highly reflective detector housing and custom phonon mask design to maximize light collection by the silicon high-voltage detector.
 
\end{abstract}

\end{frontmatter}

\section{Introduction}
\label{intro}

Cryogenic detectors have been used in a variety of rare event searches such as Low Mass Dark Matter (LMDM) search \cite{DMoverview}, Coherent Elastic Neutrino Nucleus Scattering (CE$\nu$NS) \cite{CENSFreedman1, CENS2} and Neutrino-less Double Beta Decay (NDBD) \cite{NDBD} experiments. These experiments look for a nuclear or electron recoil when a WIMP (Weakly Interacting Massive Particle) or a neutrino elastically scatters within the detector. Detection is based on the measurement of basically three types of signals: phonons (heat), scintillation (light) and ionization (charge) \cite{directdetection} caused by the recoil energy transfer. Simultaneous measurement of phonons and charges or phonons and light enhances the capability of these experiments by enabling one to discriminate between different incident particles \cite{CRESST, NUCLEUS}. The difference in the light yields for electron recoil (ER) and nuclear recoils (NR) can be used to identify these particles as well as to reject background events. The detection of CE$\nu$NS and LMDM requires detectors of large-mass and low-energy thresholds with the ability to discriminate NR and ER events. Among the cryogenic scintillation detectors, sapphire can be a very good option as it can provide low energy thresholds as well as particle identification capability at lower recoil energies. A newly developed sapphire (Al$_{2}$O$_{3}$) detector instrumented with Transition-Edge-Sensors (TES) operating at millikelvin temperatures at Texas A$\&$M University shows promising results for the detection of low recoil energy events. Because the composite detector material contains Al and O that have lower atomic masses (26 and 16 respectively) than the Si or Ge in semiconductor detectors, the resulting recoil energy from a particle interaction will be larger thereby improving the sensitivity to such interactions. In addition, because the sapphire substrate and the TES both contain Al, the phonon transmission between them is much better than for Si or Ge substrates. The results of the sapphire detector from the test facility shows a baseline resolution of 28.4 $\pm$ 0.4 eV using only the phonon channel with no bias voltage applied across the detector \cite{sapphire}. In addition to phonons, a small amount of the deposited energy in the detector is emitted as scintillation light \cite{lightdetect}. To detect such small amounts of light would require a very sensitive detector for efficient background discrimination at lower recoil energies.

To address the issue of detecting the scintillation light, our sapphire detector is paired with a phonon-mediated Si detector. To improve the threshold of the Si detector, one can exploit the Neganov Trofimov Luke (NTL) \cite{NTL, Luke} effect in which the detector is operated in a high voltage (HV) mode with an applied bias voltage up to 240V \cite{vijay}. In this paper, we show the results of the Sapphire-Silicon detector system to detect phonons as well as scintillation light. The expected linearity of the amplified photon signal with the applied voltage is shown.

\section{Combined phonon-light measurement}
\label{detection_method} 

When photons generated by particle interactions in the sapphire detector are absorbed by the Si detector they generate phonons and electron-hole pairs in the Si detector. For example, the light yield of the sapphire detector for 60 keV photons from $^{241}$Am is about 10\% $\pm$ 3\% \cite{lightdetect}. The detected signal can be enhanced by applying a bias voltage across the Si detector. The drifting electron-hole pairs in Si create additional phonons leading to an amplification of the phonon signal. The gain due to amplification is the NTL gain, $g_\mathrm{NTL} = 1 + eV/\epsilon$ where $V$ is the applied voltage across the detector, $e$ is the electronic charge and $\epsilon$ denotes the average energy required to create an electron-hole pair in Si \cite{NTL,Luke}. The NTL gain is directly proportional to the applied voltage $V$.

\section{Experimental setup and data set}
\label{data}
\begin{figure}[h]
  \centering
  \includegraphics[width=0.8\linewidth]{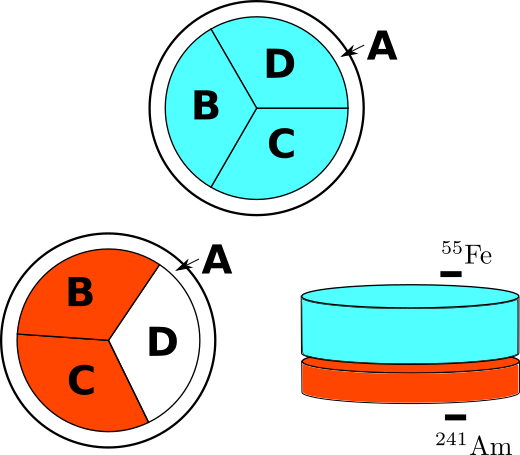}
  \caption{Schematic of the detector setup inside the dilution refrigerator where the Si HV and sapphire detectors are shown in cyan and orange colour respectively. Both detectors are paired together to form a detector module. A particle interacting with the sapphire creates two signals: phonons and photons. Phonons are read out with TES placed at the top surface of the sapphire whereas the photons generated in the sapphire are collected in the Si detector which is also equipped with TES. The TES at the surfaces of both detectors are divided into four independent readout channels. One outer channel A and three inner channels B, C and D are highlighted in the figure for both detectors. The readout channels are the three inner channels (B, C, D) for the Si detector and the two inner channels (B, C) for the sapphire detector shown in cyan and orange respectively. Two known energy sources are used for calibration, an $^{55}$Fe source placed at the top of the Si detector and an $^{241}$Am source placed at the bottom of the sapphire detector as shown in the figure.}
  \label{fig:detsetup}
\end{figure}

Both sapphire and Si detectors developed by the Texas A\&M group have a mass of 100 g. The sapphire detector has a diameter of 76 mm and a thickness of 4 mm, whereas the Si detector has a diameter of 76 mm and a thickness of 10 mm. The sapphire and the Si HV detectors are paired together as a detector module and placed inside a BluFors dilution refrigerator that utilizes cryocoolers to maintain a base temperature of $\sim$ 8 mK. The schematic of the detector module inside the refrigerator is shown in Fig.\ref{fig:detsetup} where the sapphire and Si HV detectors are shown in orange and cyan respectively. The detectors are separated by a gap of 2 mm. One face of each detector is covered with TES arrays for phonon signal readout. The sensors are divided into 4 independent groups (A, B, C, D) as shown in the figure where the readout channels are highlighted. To calibrate the Si detector an $^{55}$Fe gamma source is placed at the top of the detector surface and an $^{241}$Am source (the 5 MeV alpha particles are shielded) is placed at the bottom of the sapphire detector. The light produced in the sapphire is collected in the Si HV detector using coincidence techniques. No reflector has been used in this experimental setup. The Si HV detector is operated with a bias voltage up to 100 V to study the NTL gain of the light output from the sapphire.

\begin{figure*}[t]
  \centering  
    \includegraphics[width=\linewidth]{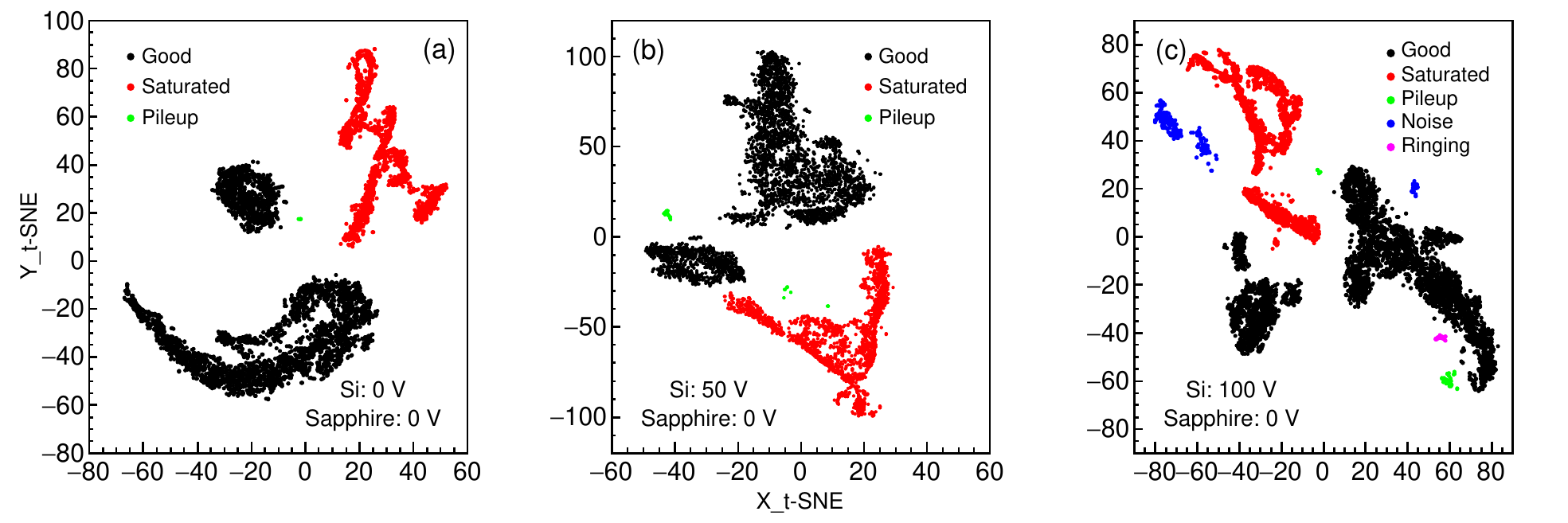}\\
    \caption{t-SNE and DBSCAN response plots where X and Y axis represent t-SNE x and t-SNE y axes. (a), (b) and (c) show the data in 2-dimensional t-SNE space after clustering with three different voltages, 0 V, 50 V and 100 V respectively. In (a) and (b), black, red and green represent good, saturated and pileup events respectively. In (c) additionally, we include typical noise and ringing events indicated by blue and pink markers respectively. Good events (black) are chosen for further data analysis for all three voltage data.}
  \label{fig:t-SNE}
\end{figure*}

\section{Analysis and Result}
\label{results}
The raw data acquired from the DAQ contains the information of each event in the form of a voltage as a function of time. Because the recorded pulses can be either signal or noise, we describe in the next section an anomaly detection (AD) technique that is used to improve the data quality during pulse filtration in the data analysis.

\begin{table*}
\centering
\caption{Description of the feature vector. \{ch\} refers to channels B and C.}
\label{tab_fe}
{\small%
\begin{tabular}{ccp{0.25\linewidth}|ccp{0.25\linewidth}}
\hline
Sl. No. 	&	 Feature Name    	&	 Description                                                    &	Sl. No. 	&	 Feature Name    	&	 Description                                                    \\ \hline
1	&	 Pre-pulse SD \{ch\}	&	 Standard deviation of the pulse amplitude for the first 500 $\mu$s for Sapphire \{ch\} 	&	7	&	 Rise Time \{ch\}   	&	 Time to go from 10$\%$ to 90$\%$ of the peak amplitude for Sapphire \{ch\}     \\
2	&	 Post-pulse SD \{ch\}	&	 Standard deviation of the pulse amplitude for the last 256 $\mu$s for Sapphire \{ch\} &	8	&	 Fall Time \{ch\}   	&	 Time to go from 90$\%$ to 10$\%$ of the peak amplitude for Sapphire \{ch\}     \\
3	&	 Max Time \{ch\}    	&	 Time of the peak amplitude for Sapphire \{ch\}                                &	9	&	 FW10M \{ch\}        	&	 Full width at 10$\%$ of the amplitude for Sapphire \{ch\}                \\
4	&	 Min Time \{ch\}   	&	 Time of the minimum amplitude for Sapphire \{ch\}                                &	10	&	 FWHM \{ch\}        	&	 Full width at 50$\%$ of the peak amplitude for Sapphire \{ch\}                 \\
5	&	 Peak \{ch\}        	&	 Peak amplitude for Sapphire \{ch\}                                            &	11	&	 FW90M \{ch\}        	&	 Full width at 90$\%$ of the peak amplitude for Sapphire \{ch\}                 \\
6	&	 Max Tail \{ch\}   	&	 Maximum value for the last 256 $\mu$s for Sapphire \{ch\}                    &	12	&	 Max SD          	&	 Standard deviation of time of peak amplitudes among channel B and C   \\ \hline
\end{tabular}}
\end{table*}

\begin{figure*}[t]
\centering
\begin{minipage}{0.33\linewidth}
    \includegraphics[width=0.99\linewidth]{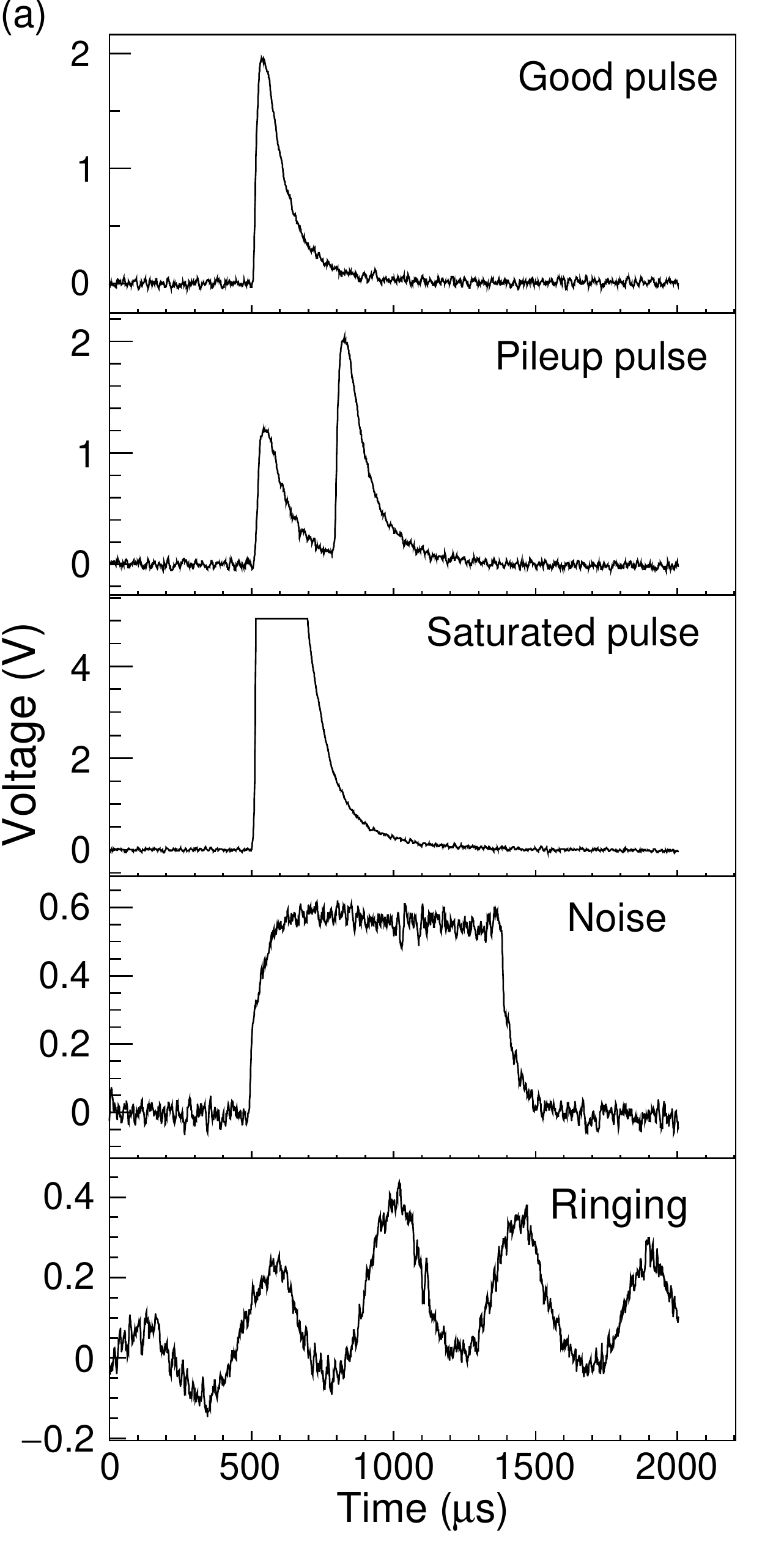}
\end{minipage}   
\begin{minipage}{0.64\linewidth}
\includegraphics[width=0.5\linewidth]{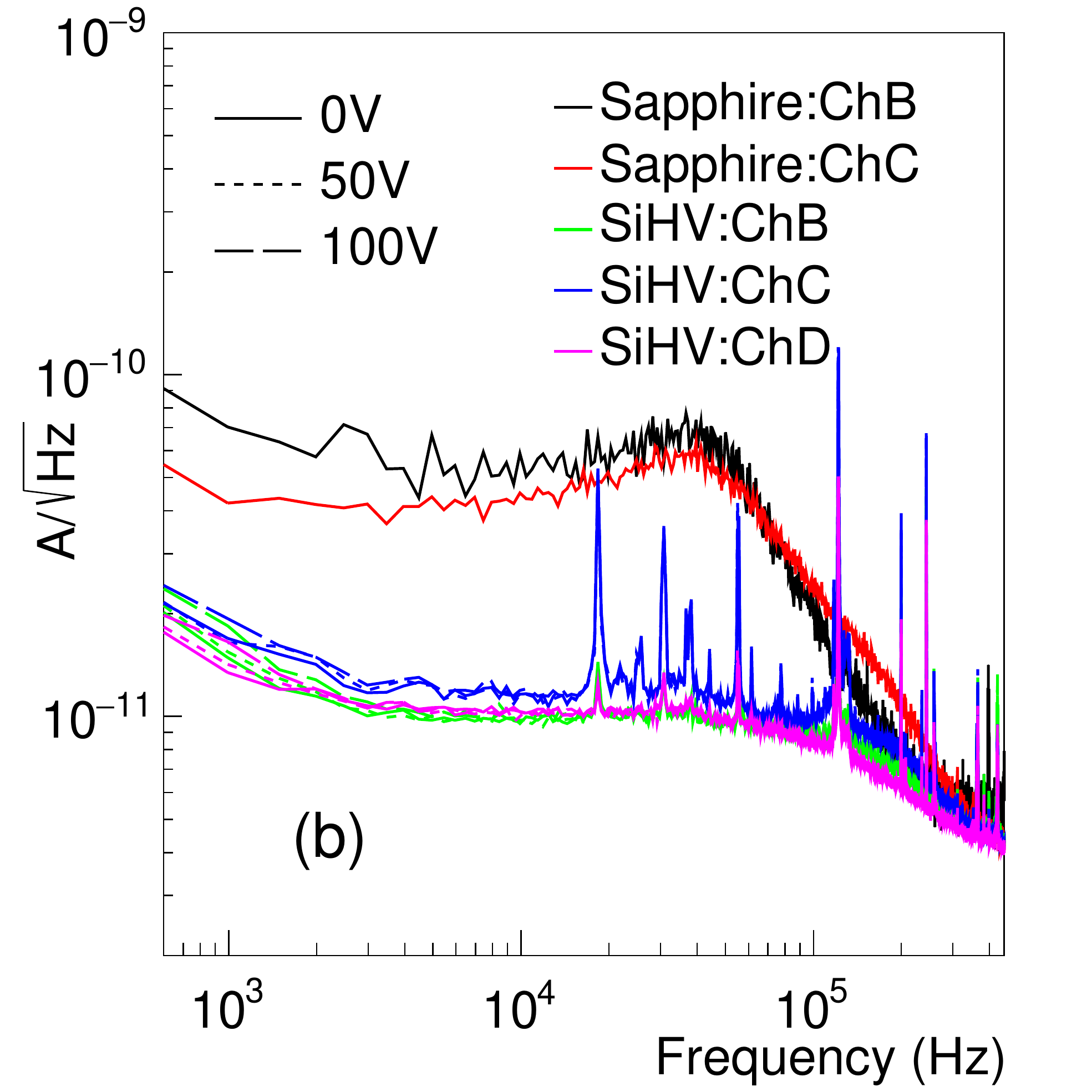}
\includegraphics[width=0.5\linewidth]{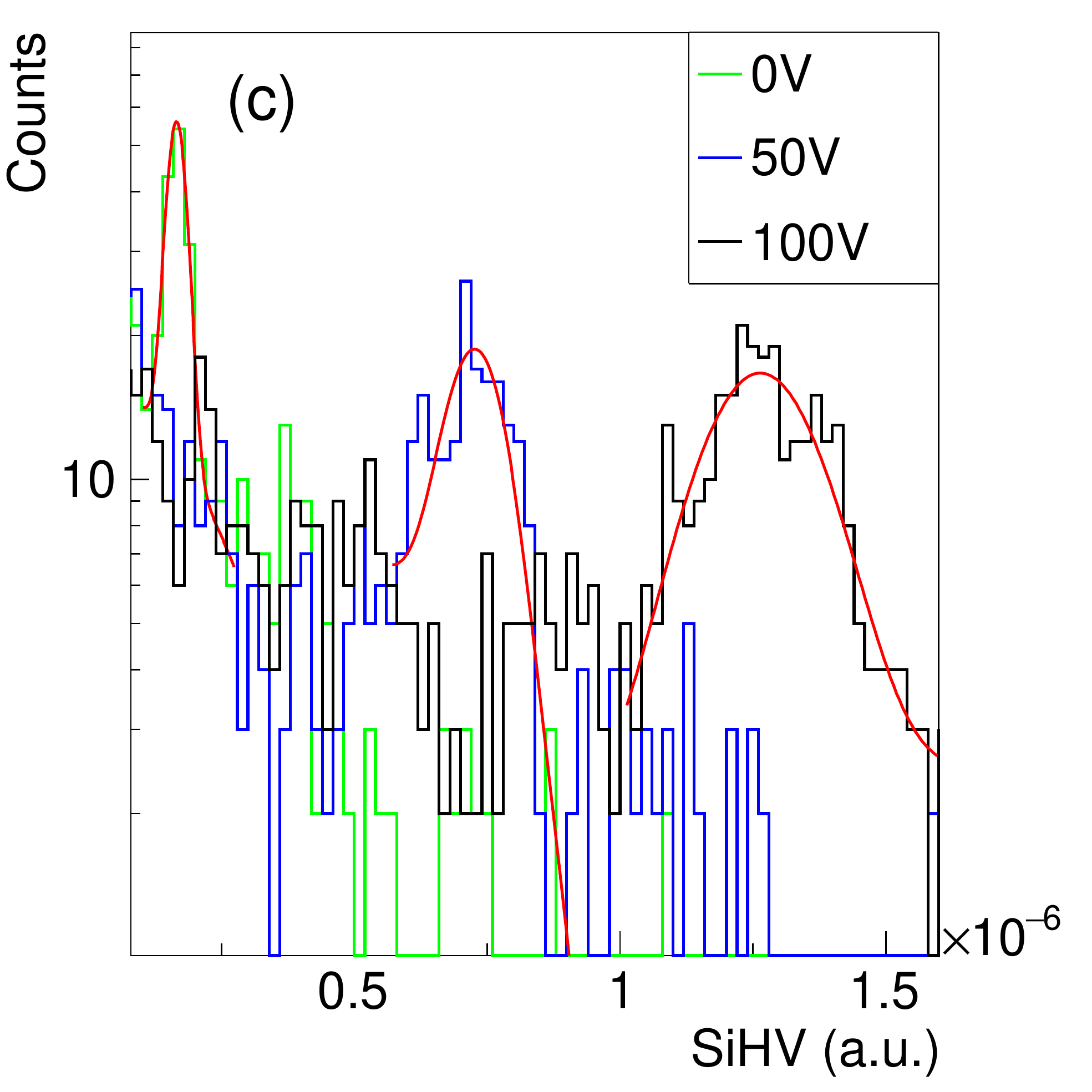}\\
  \includegraphics[width=0.5\linewidth]{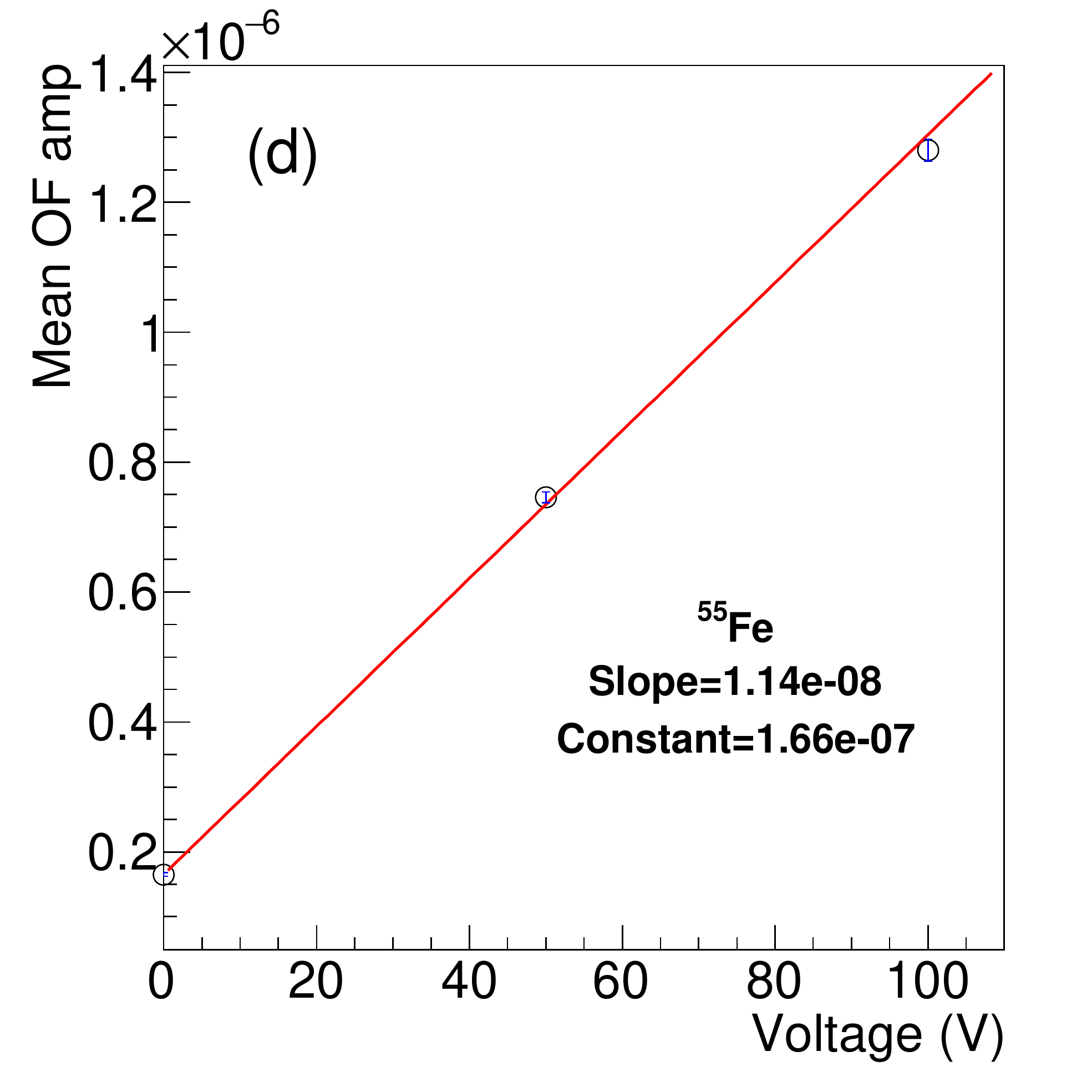}
  \includegraphics[width=0.5\linewidth]{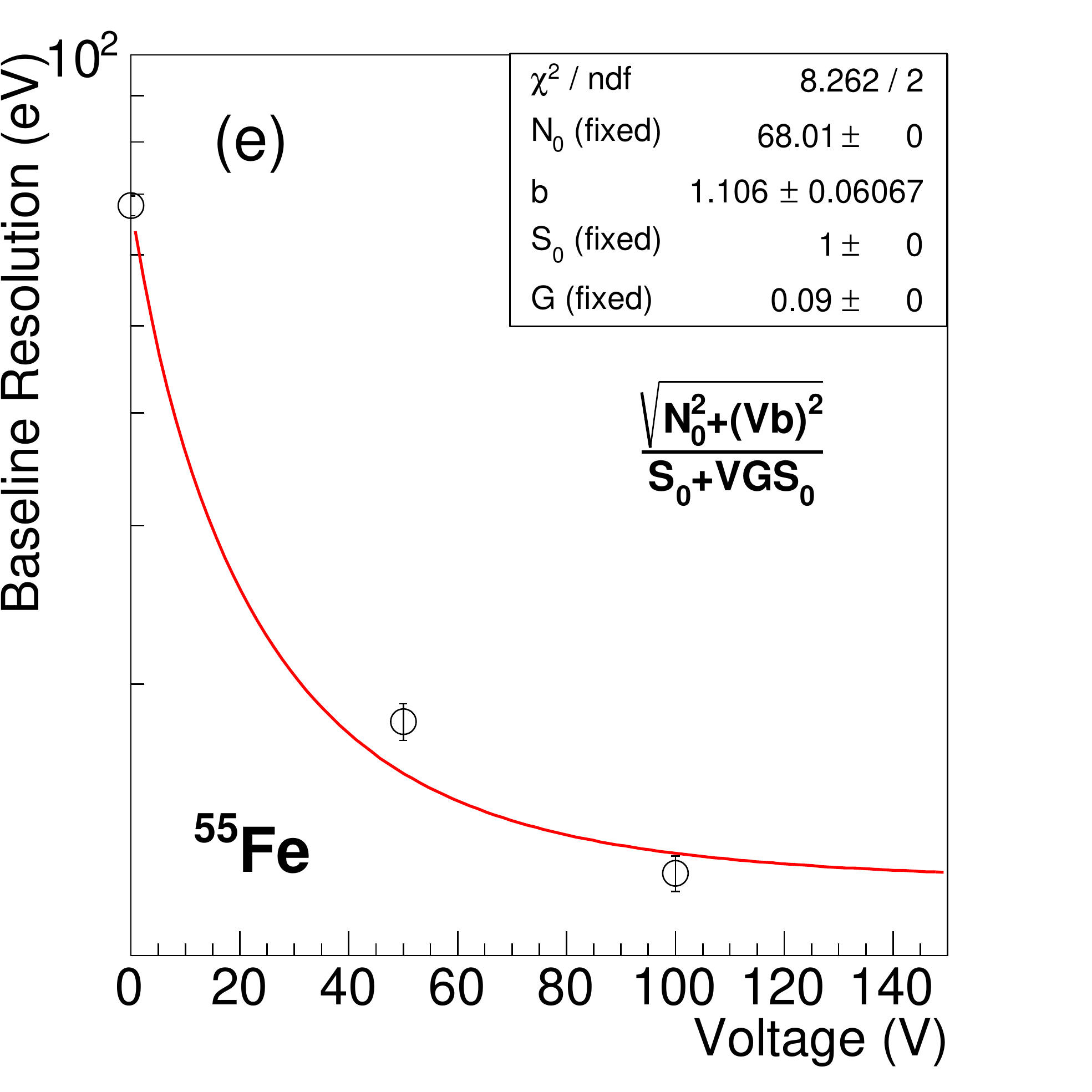}
\end{minipage}
\caption{(a) Examples of a typical good pulse, pileup pulse, saturated pulse, noise, and ringing pulse are shown for the 100 V data set. Voltage is plotted as a function of time in microseconds. (b) Phonon noise power spectral density (PSD) where the current amplitude in $A/\sqrt{\mathrm{Hz}}$ is plotted as a function of frequency in Hz. The PSD for the two sapphire channels at 0 V is shown in the black and red solid curves. The PSD for the three channels in the Si HV detector are shown in green, blue, and pink with solid, dotted and dashed lines corresponding to 0 V, 50 V, and 100 V respectively. (c) 6 keV peaks from $^{55}$Fe source in the Si HV detector at 0 V, 50V and 100V (green, blue and black lines respectively) with gaussian fits (red lines). (d) The mean of the peak in OF unit is plotted with the applied three voltages, 0 V, 50 V, and 100 V. A linearity in amplifying the phonon signal with the applied voltages is seen in the experimental data. (e) Baseline resolution in eV is plotted with applied voltages. Statistical errors are shown with the data points. The lowest value of baseline resolution is 12 eV at 100 V which indicates that the S/N ratio is improving with the application of higher voltages.}
\label{fig:SiHV}
\end{figure*}

\subsection{Anomaly detection: t-SNE and DBSCAN}
Anomaly detection is the identification of events that differ significantly from the majority of the events in the data \cite{Zimek2017}. In a dataset, AD tries to differentiate between the events on the basis of their similarity/dissimilarity with other events. In this work, we have used a combination of two algorithms to cluster our data, t-SNE \cite{tsne} and DBSCAN \cite{dbscan} that are used in a sequential manner where the output from t-SNE is used to feed into the DBSCAN algorithm. t-SNE stands for t-distributed stochastic neighborhood embedding. It is used to embed the data from a higher dimensional space to a lower dimension space while keeping similar events together. It is an unsupervised machine learning (ML) technique that understands the overall trends and patterns without any prior knowledge of the data. Density-based spatial clustering of application with noise (DBSCAN) is a clustering method that uses spatial information and groups data based on the relative distance between them. We refer to these groups as clusters. DBSCAN was used instead of other clustering algorithms, as it does not need a predefined number of clusters into which the data has to be divided and it is not constrained by the shape of the clusters. 

In our analysis, the shape of the pulses from channels B and C of the sapphire detector is characterized by a 23-dimensional vector that we call a feature vector. The description of the feature vector is tabulated in Table \ref{tab_fe}. An event is any interaction in the detector that triggers a readout. 
We have only used output from the sapphire TES to characterize the data because the trigger was on the sapphire detector and the pulses from the sapphire detector had a much steeper rise time and fall time than pulses from the Si TES. Using the feature vector we represent each event in 23 dimensions. The Pre-pulse and Post-pulse standard deviation (SD) can help t-SNE to learn about the noisiness of the pulse, while rise and fall times and widths at different amplitudes help to capture the shape of the pulse. The times of maximum and minimum amplitudes depend on the trigger and any anomalous event that might deviate from the dataset on these variables. The last variable, Max SD, characterizes the range of time the detector took to detect the pulse in different channels. We expect the times to be slightly different but not very far away from each other. This information is used to embed the 23-dimension feature vector into a 2-dimensional space using t-SNE. After this, we cluster the data in the 2-dimensional space using DBSCAN. Fig.\ref{fig:t-SNE} shows different clusters in the 2-dimensional t-SNE space using the data from  three different bias voltages across the Si detector. We now examine the different clusters that result from DBSCAN and group clusters that correspond to saturated pulses, good pulses, piled-up pulses, ringing, and noise. There can be multiple clusters with good pulses, for example at 0 V there were two clusters with good pulses. These clusters contained pulses with different amplitude ranges. We did not need such classification and hence labeled both clusters as good pulses. These clusters are shown in different colors in Fig.\ref{fig:t-SNE}. Good pulses are shown in black, saturated pulses in red, pile-ups in green, noise in blue, and ringing in magenta. Examples of a typical good pulse, pile-up pulse, saturated pulse, noise, and ringing are shown in Fig.\ref{fig:SiHV}(a). Events from the cluster which represent good pulses are filtered out for further analysis.

Traditionally the way to filter bad pulses from good pulses is by making use of a pulse template. One generates a pure pulse template from the unfiltered data itself. This template is fitted on the pulse and a $\chi^{2}$ cut is used to filter good pulses. In an unsupervised machine learning algorithm such as the one proposed above, we do not need to provide any information for the filtration other than the dataset itself. Our feature vector tries to understand the shape of the pulses and represents them as a vector and t-SNE converts this to a 2-dimensional space while keeping events similar in shape together. DBSCAN is then used to group these events based on their density. However, in the pulse template fit method filtration is based on the template which is created by a human based on his judgment from an unfiltered dataset. We measured the efficiency and purity of our proposed method by creating a labeled dataset of images. We extracted nine events with good pulses, nine with saturated pulses, nine with piled-up pulses, three with noise pulses, and three with ringing pulses from our dataset. Then a new dataset with 6008 events was created by linearly combining these events within their types and labeled. There were 1993 good pulses in the new dataset. After applying t-SNE and DBSCAN to the dataset we obtained twenty clusters. Clusters that contained mostly good pulses were filtered out. The purity of the filtered pulses was 98.1$\%$. Out of 1993 good pulses, 1990 were filtered out and we missed 3 pulses that give us an efficiency of 99.8$\%$. This accuracy was obtained on a labeled subset of the full analysis dataset. We expect a similarly high accuracy when applying this pulse selection method on the full dataset.

\begin{figure*}[t]
   \centering
\begin{subfigure}[t]{0.33\linewidth}
  \centering
  \includegraphics[width=\linewidth]{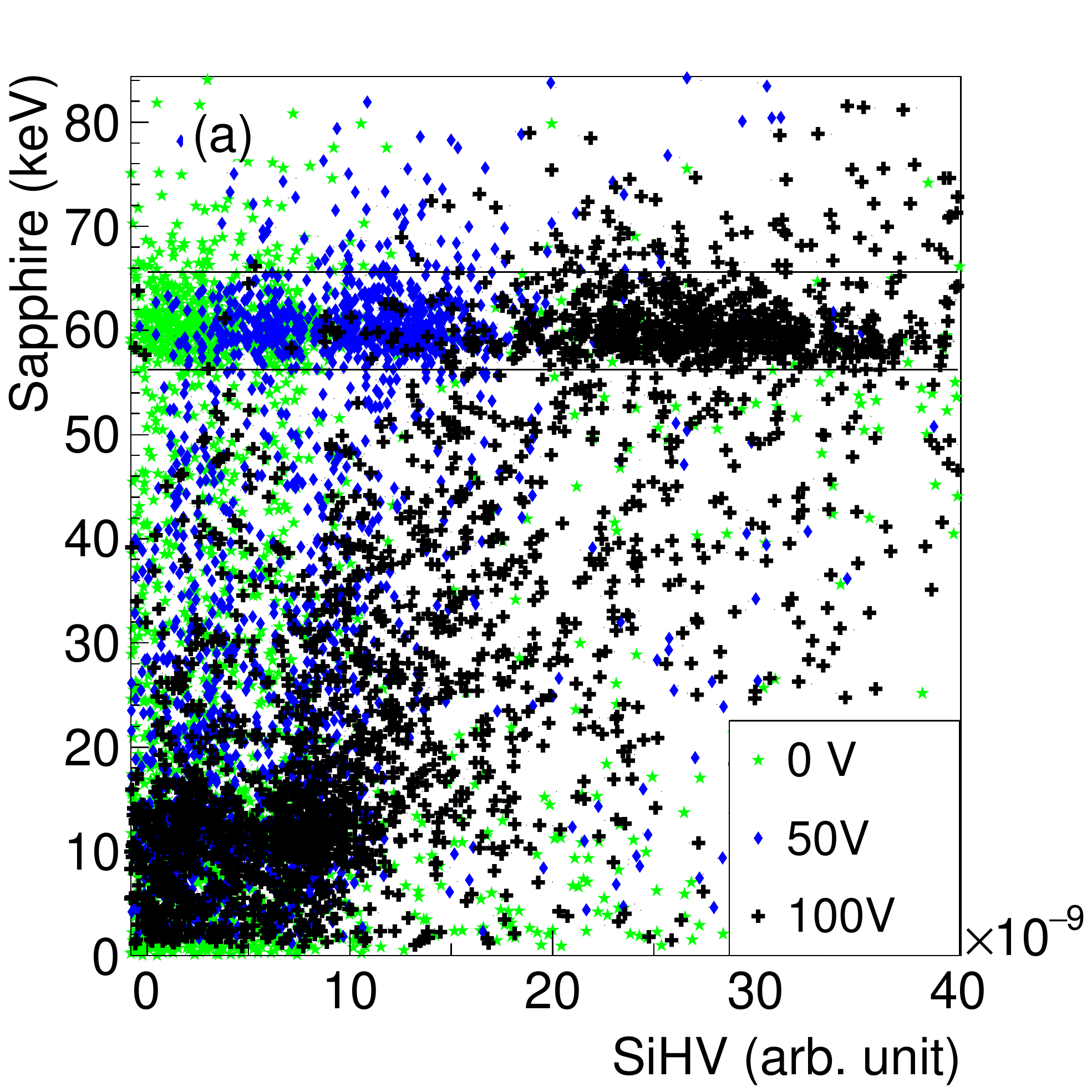}
\end{subfigure}
\begin{subfigure}[t]{0.33\linewidth}
  \centering
  \includegraphics[width=\linewidth]{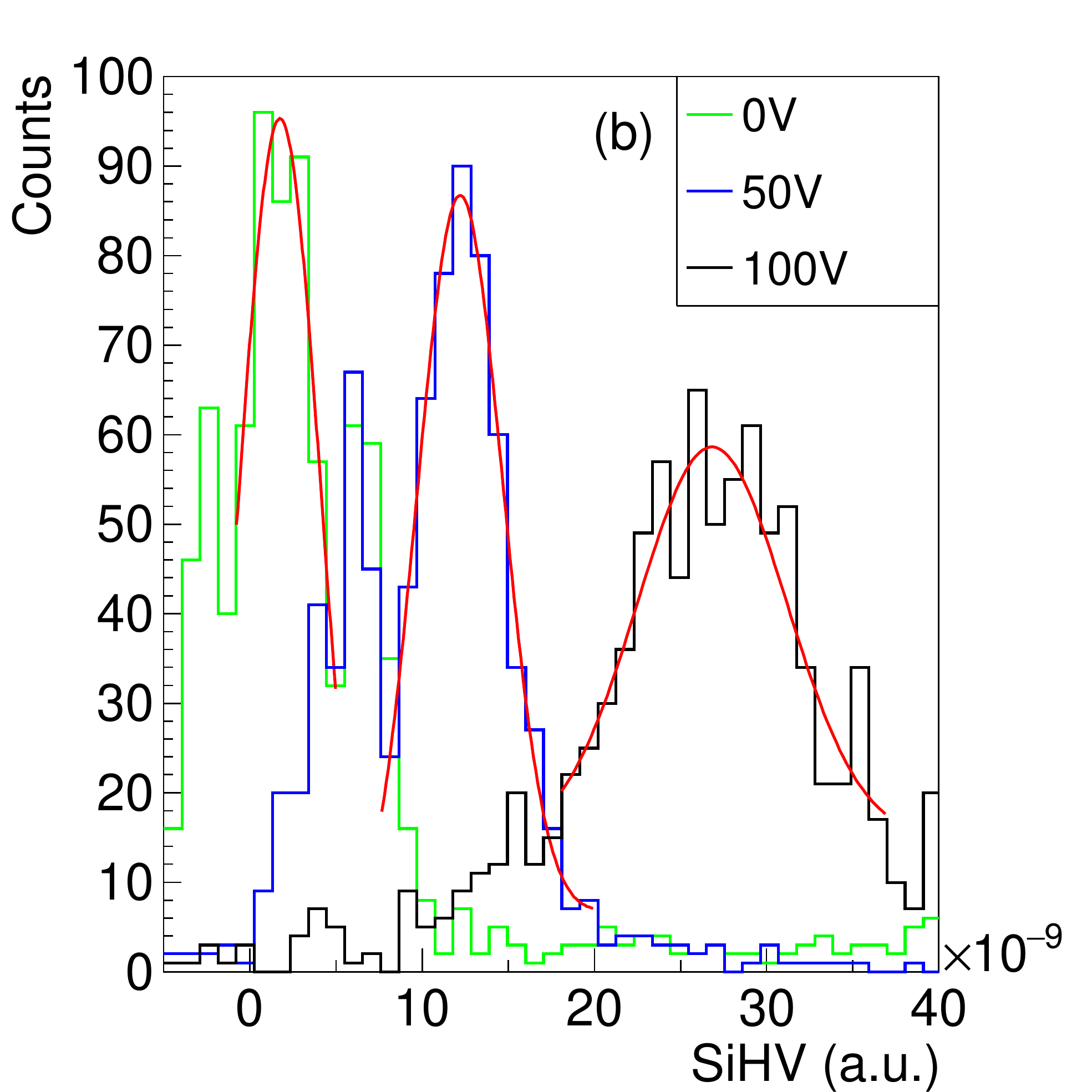}
\end{subfigure}
\begin{subfigure}[t]{0.33\linewidth}
  \centering
  \includegraphics[width=\linewidth]{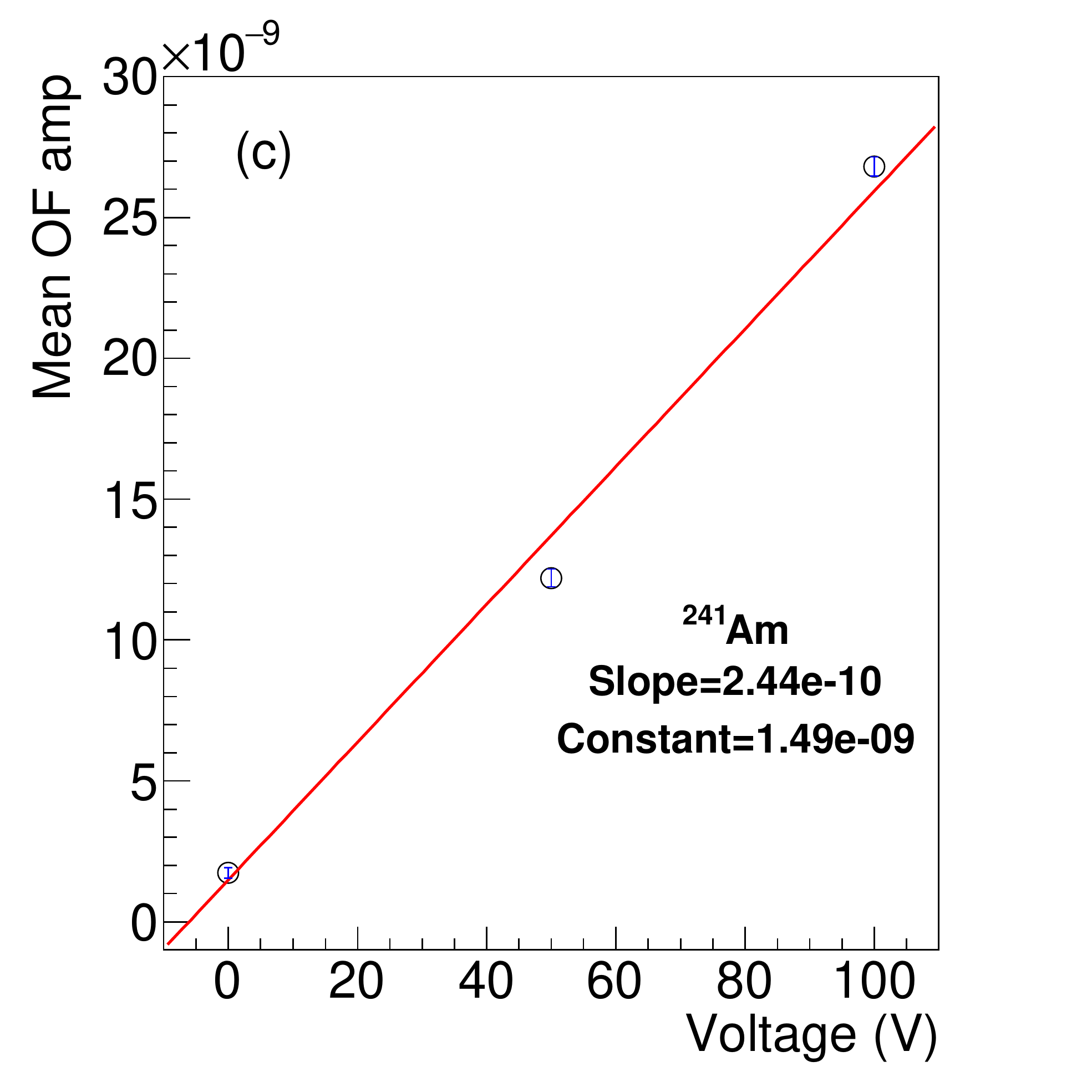}
\end{subfigure}
\caption{(a) Correlation plot between phonon energy measured in sapphire (on the Y-axis) plotted with the light output measured in Si HV detector (on the X-axis) at three different voltages. The light output of 60 keV events from an $^{241}$Am source in the sapphire detector is amplified in the Si HV detector. The projected 60 keV events (shown in black lines) in the Si HV detector showing the amplified light at different voltages are shown in figure (b). (c) shows the linearity of the amplified light as a function of voltage. Statistical errors associated with the data points are shown.}
\label{fig:sapphireSiHV}
\end{figure*}

\subsection{Processing raw data after filtering: Optimal filter (OF) method}
We have used pulse template fitting algorithm known as Optimal filter (OF) algorithm \cite{OFmethod} to extract energy information from the raw traces of the filtered data set. Noise power spectral density (PSD) with few noise pulses is calculated from the dataset taken with random triggers. The OF algorithm is then applied to the raw data set where the pulses are fitted with the template in the frequency domain to determine the pulse amplitude from the best $\chi^{2}$ fit values. The amplitude measured in the OF method is directly proportional to the energy of the pulses. The OF amplitude distribution can then be calibrated using known energy sources. Similarly, the random triggered data set which is the measure of the noise can be used to calculate the baseline resolution of the detectors.

\subsection{Calibration of Si detector}
Si HV detector is calibrated with 6 keV gammas from the $^{55}$Fe source. The data was taken with the bias voltage at 0 V, 50 and 100 V. Fig.\ref{fig:SiHV}(b) shows the noise performance of the sapphire channels and the Si channels when the Si HV detector was biased at 0 V, 50 V and 100 V. The noise in the Si channels does appear to increase with increasing bias voltage applied to the Si HV detector. Fig.\ref{fig:SiHV}(c) shows the 6 keV peak from the $^{55}$Fe source measured at the three bias voltages and fitted with a gaussian. The mean value of the distribution is plotted at different voltages and shown in Fig.\ref{fig:SiHV}(d). As expected the NTL amplification of the 6 keV events is linearly proportional to the applied voltages as shown by the straight line in the figure. The baseline resolution was calculated using random triggered data for all three voltages. Fig.\ref{fig:SiHV}(e) shows the baseline resolution as a function of the voltage where the lowest baseline resolution achieved is 12 eV.

\subsection{Amplification of light in Si detector}
\label{amplify}
In this section, we discuss the amplification of the signal in the Si HV detector due to the scintillation light from the sapphire detector by examining only coincident events in the sapphire and Si detectors. An $^{241}$Am source is used to illuminate the sapphire detector with 60 keV gammas.
When a 60 keV gamma is absorbed in the sapphire, both phonons and scintillation light are created. On average only 10\% of the energy of the incident gamma is converted into light \cite{lightdetect}. The phonons are measured by the TES on the sapphire detector whereas only the photons that reach the Si HV detector are detected by the TES on the Si detector. A Monte Carlo simulation to estimate the geometric efficiency in collecting the photons by the Si HV detector indicated that only 40\% of the photons reach the Si detector. Moreover, the high voltage phonon mask on the Si HV detector facing the sapphire detector covers 90\% of the surface, further reducing the light from entering the Si detector. Finally given the reflectivity of Si to be $\sim$ 56\% \cite{GREEN20081305, Optical_prop_Si}, we expect that the effective light collection to be only $\sim$ 2\%. The scintillation photons of $\sim$6 keV (10 $\%$ of 60 keV gammas) total energy (after undergoing losses from solid angle acceptance and absorption) will hit the Si detector and create phonon signals in Si. The amplification of the light signal in the Si detector is observed through coincidence events with the sapphire detector. The correlation between the sapphire and Si detectors is shown in Fig.\ref{fig:sapphireSiHV}(a) where we see that light output from the 60 keV events in sapphire gets amplified in the Si detector with the applied voltages. Fig.\ref{fig:sapphireSiHV}(b) shows the projection of 60 keV events in the Si detector which demonstrate the amplification of the light at different voltages. The measured light in the Si HV detector is converted to energy by multiplying the calibration factor of the detector. The amount of light we measure in our detector is 2\% which matches well with the expected number after considering all the losses. To show the linear relationship we have plotted the mean of the projected event distribution with the respective voltages. Fig.\ref{fig:sapphireSiHV}(c) shows the linearity of the amplified light output in the Si detector. 

\section{Conclusion and outlook}
\label{conclusion and outlook}
In this paper, we demonstrate the simultaneous detection of phonons and scintillation light in a system composed of a sapphire and Si HV detector. Despite using any reflector in the experiment, we are able to measure 2\% of the total light produced in the sapphire which agrees well with the expected value after considering all the losses. The lowest energy threshold achieved in the Si HV detector is 36 eV (3 times the baseline resolution) at 100 V. As the Si HV detector can be operated at voltages up to 240 V, our future work aims to further lower the energy threshold by applying higher voltages. We will also focus on improving the light collection efficiency by having a reflective detector housing and a Si HV detector with a custom phonon mask design that covers less area of the Si HV detector surface for maximum light collection efficiency. We can conclude that this type of detector system can be used in the search for low-mass dark matter and CE$\nu$NS where low energy recoils can be detected with active background discrimination.

\section*{Acknowledgement} 
This work was supported by the U. S. Department of Energy (DOE) under Grant Nos DE-SC0020097, DE-SC0018981, DE-SC0017859, and DE-SC0021051. We acknowledge the seed funding provided by the Mitchell Institute for early conceptual and prototype development. We would like to acknowledge the support of DAE-India through the project Research in Basic Sciences - Dark Matter and SERB-DST-India through the J. C. Bose Fellowship.


\bibliography{references}

\end{document}